\documentclass[aps,prb,reprint,longbibliography,superscriptaddress]{revtex4-2}
\usepackage{mathrsfs}
\usepackage{amsmath,gensymb}
\usepackage{amsfonts}
\usepackage{amssymb}
\usepackage{amsthm}
\usepackage{graphicx}
\usepackage{natbib}
\usepackage{color}
\usepackage{hyperref}
\usepackage{bm}
\usepackage[caption=false]{subfig}
\usepackage{verbatim}

\newcommand{\ket}[1]{\left| {#1} \right\rangle}
\newcommand{\bra}[1]{\left\langle {#1} \right|}

\begin{document}
\title{Tunable anisotropic quantum Rabi model via a magnon--spin-qubit ensemble}

\author{Ida C. Skogvoll}
\affiliation{Center for Quantum Spintronics, Department of Physics, Norwegian University of Science and Technology, NO-7491 Trondheim, Norway}

\author{Jonas Lidal}
\affiliation{Center for Quantum Spintronics, Department of Physics, Norwegian University of Science and Technology, NO-7491 Trondheim, Norway}

\author{Jeroen Danon}
\affiliation{Center for Quantum Spintronics, Department of Physics, Norwegian University of Science and Technology, NO-7491 Trondheim, Norway}

\author{Akashdeep Kamra}
\email{akashdeep.kamra@uam.es}
\affiliation{Condensed Matter Physics Center (IFIMAC) and Departamento de F\'{i}sica Te\'{o}rica de la Materia Condensada, Universidad Aut\'{o}noma de Madrid, E-28049 Madrid, Spain}
\affiliation{Center for Quantum Spintronics, Department of Physics, Norwegian University of Science and Technology, NO-7491 Trondheim, Norway}

\begin{abstract}
The ongoing rapid progress towards quantum technologies relies on new hybrid platforms optimized for specific quantum computation and communication tasks, and researchers are striving to achieve such platforms. We study theoretically a spin qubit exchange-coupled to an anisotropic ferromagnet that hosts magnons with a controllable degree of intrinsic squeezing. We find this system to physically realize the quantum Rabi model from the isotropic to the Jaynes-Cummings limit with coupling strengths that can reach the deep-strong regime. We demonstrate that the composite nature of the squeezed magnon enables concurrent excitation of three spin qubits coupled to the same magnet. Thus, three-qubit Greenberger-Horne-Zeilinger and related states needed for implementing Shor’s quantum error-correction code can be robustly generated. Our analysis highlights some unique advantages offered by this hybrid platform, and we hope that it will motivate corresponding experimental efforts.  
\end{abstract}

\maketitle


\section{Introduction}\label{sec:intro}

A bosonic mode interacting with a two-level system constitutes the paradigmatic quantum Rabi model (QRM) employed in understanding light-matter interaction~\cite{Rabi1936,Rabi1937}. The recent theoretical discovery of its integrability~\cite{Braak2011} and increasing coupling strengths realized in experiments have brought the QRM into a sharp focus~\cite{Xie2014,Xie2017}. It also models a qubit interacting with an electromagnetic mode, a key ingredient for quantum communication and distant qubit-qubit coupling~\cite{Wehner2018,Fukami2021,Awschalom2021,Burkard2020}. Thus, the ongoing quantum information revolution~\cite{Wehner2018,Laucht2021} capitalizes heavily on the advancements in physically realizing and theoretically understanding the QRM. In particular, larger coupling strengths are advantageous for faster gate operations on qubits, racing against imminent decoherence. Generating squeezed states of the bosonic mode~\cite{Gerry2004,Walls1983}, typically light, via parametric amplification has emerged as a nonequilibrium means of strengthening this coupling {and achieving various entangled states~\cite{Qin2018,Leroux2018,Lu2015,Chen2019,Chen2021,Burd2021}.} Other related methods~\cite{Wang2019,Sanchez2020} that exploit drives to control, for example, the QRM anisotropy~\cite{Xie2014} have also been proposed.

Contemporary digital electronics relies heavily on the very-large-scale integration of the same silicon-based circuits. In sharp contrast, emerging quantum information technologies benefit from multiple physical realizations of qubits and their interconnects in order to choose the best platform for implementing a specific task or computation~\cite{Wehner2018,Russ2017,Chatterjee2021,Vandersypen2019,Lachance-Quirion2019,Awschalom2021}. Fault-tolerant quantum computing, either via less error-prone qubits~\cite{Nayak2008} or via implementation of quantum error correction~\cite{Shor1995,Terhal2020,Gottesman2001}, is widely seen as the path forward. A paradigmatic error correction code~\cite{Shor1995} put forth by Shor requires encoding one logical qubit into 9 physical qubits and generating 3-qubit Greenberger-Horne-Zeilinger (GHZ)~\cite{Greenberger2007} and related states. A continuous-variable analog of this code employing squeezed states of light has been experimentally demonstrated~\cite{Aoki2009}. This has spurred fresh hopes of fault-tolerant quantum computing and demonstrated the bosonic modes as more than just interconnects for qubits.   

In our discussion above, we have encountered squeezed states of light in multiple contexts. These nonequilibrium states, bearing widespread applications from metrology~\cite{Schnabel2017} to quantum teleportation~\cite{Ou1992,Milburn1999B}, decay with time. In contrast, the bosonic normal modes - magnons - in anisotropic ferromagnets were recently shown to be squeezed~\cite{Kamra2016A} and embody various quantum features inherent to such squeezed states~\cite{Gerry2004,Kamra2020,Zou2020,Sharma2021}. Being equilibrium in nature, these are also somewhat different from light and require care when making comparisons. This calls for examining ways in which we can exploit the robust equilibrium-squeezed nature of magnons in addressing challenges facing emerging quantum technologies~\cite{Wang2020,Lachance-Quirion2019,Tabuchi2015}. The spin qubit~\cite{Vandersypen2019,Loss1998,Chatterjee2021} becomes the perfect partner because of its potential silicon-based nature, feasibility of a strong exchange-coupling to the magnet, reliance on a mature fabrication technology and so on. 

Here, we theoretically study a ferromagnet exchange-coupled to a spin qubit. We find the ensuing magnon--spin-qubit ensemble to combine various complementary advantages mentioned above into one promising platform. We show that this system realizes an ideal Jaynes-Cummings model, enabled by spin conservation in the system that forbids the counter rotating terms (CRTs) by symmetry. Allowing anisotropy in the magnet, the squeezed-magnon~\cite{Kamra2016A,Kamra2020} becomes the normal mode giving rise to nonzero and controllable CRTs. The squeezed nature of the magnon further leads to an enhancement in the coupling strength, without the need for a nonequilibrium drive. Considering three spin qubits coupled to the same ferromagnet, we theoretically demonstrate the simultaneous and resonant excitation of the three qubits via a single squeezed-magnon. Thus, the system enables a robust means to generate the entangled 3-qubit GHZ and related states that underlie Shor's error correction code~\cite{Shor1995}. The magnon--spin-qubit ensemble offers an optimal platform for realizing the QRM with large coupling strengths and implementing fault-tolerant quantum computing protocols.


\section{1 magnonic mode coupled to 1 qubit}\label{sec:1mag1qubit}

\begin{figure}[tb]
	\begin{center}
		\includegraphics[width=85mm]{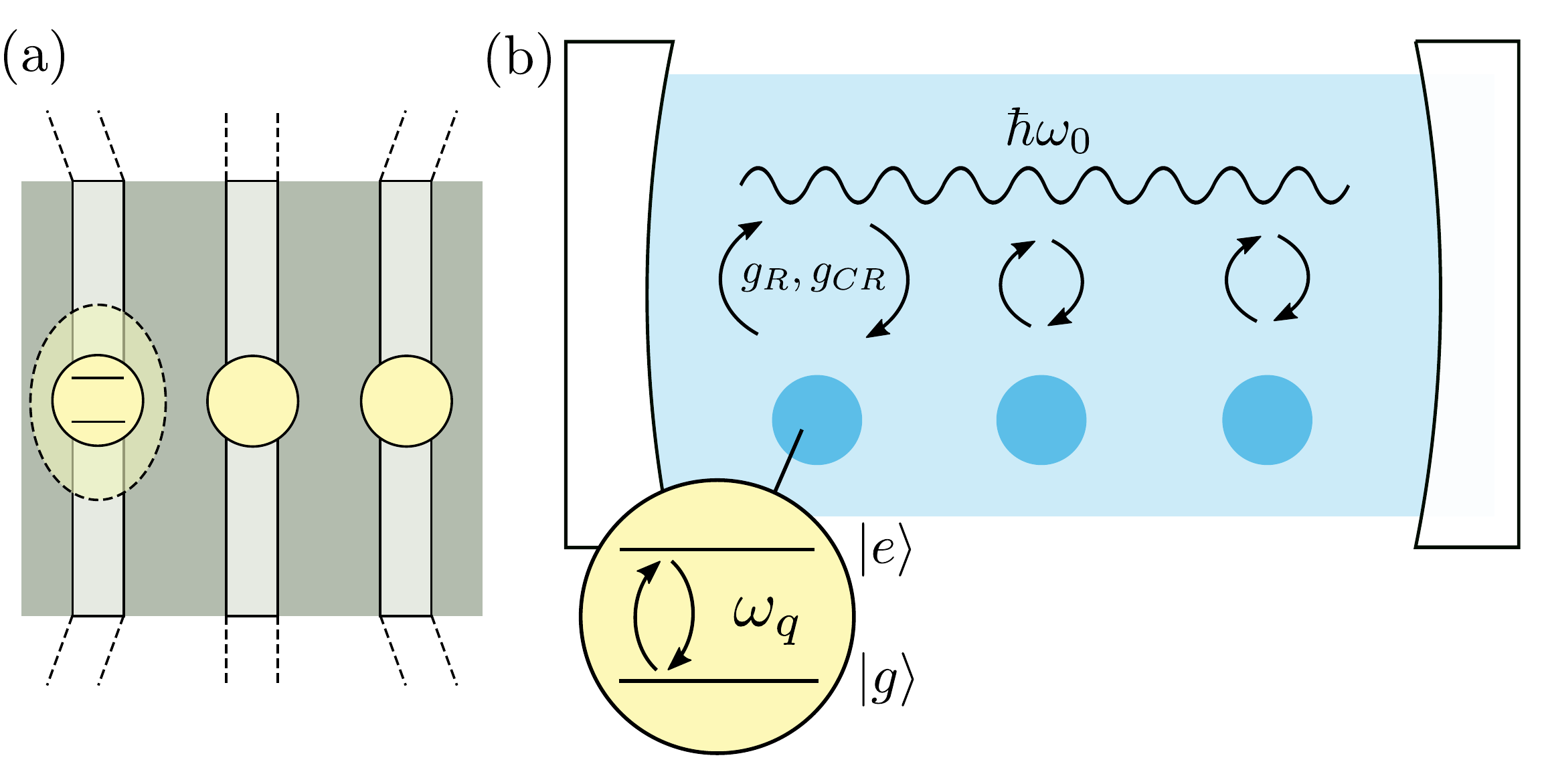}
		\caption{Schematic depiction of 3 spin qubits exchange-coupled to 1 magnon mode. (a) Semiconducting wires hosting the localized electronic states that constitute the spin qubit are deposited on top of a thin insulating ferromagnet layer. A direct contact enables strong interfacial exchange coupling. (b) The corresponding anisotropic QRM. Three qubits interact with a single magnonic mode via controllably strong rotating ($g_R$) and counter-rotating ($g_{CR}$) terms [Eq.~\eqref{eq:hint2}].}
		\label{fig:1}
	\end{center}
\end{figure}

We consider a thin film of an insulating ferromagnet that acts as a magnonic cavity. Considering an applied magnetic field $H_0 \hat{\pmb{z}}$, the ferromagnetic Hamiltonian is expressed as~\cite{Akhiezer1968}:
\begin{align}
\tilde{H}_{\mathrm{F}} & = - J  \sum_{\left \langle i,j \right \rangle}   \tilde{\pmb{S}}_i \cdot \tilde{\pmb{S}}_j + |\gamma| \mu_0 H_0 \sum_{i} \tilde{S}_{iz},
\end{align}
where $J~(>0)$ parametrizes ferromagnetic exchange between the nearest neighbors, $\gamma~(< 0)$ is the gyromagnetic ratio, and $\tilde{\pmb{S}}_i$ denotes the spin operator at position $i$. We set $\hbar = 1$ throughout and identify operators with an overhead tilde. A detailed derivation of the system Hamiltonian is presented in {appendix \ref{sec:ham}. We discuss the key steps and their physical implications in the main text. Due to the Zeeman energy, the ferromagnet has all its spins pointing along $- \hat{\pmb{z}}$ in its ground state.} Employing Holstein-Primakoff transformations~\cite{Holstein1940} and switching to Fourier space, the ferromagnetic Hamiltonian is written in terms of spin-1 magnons~\cite{Kittel1963}:
\begin{align}
\tilde{H}_{\mathrm{F}} & = \mathrm{const.} + \sum_{\pmb{k}} \left(\omega_0 + c_l J S a^2 k^2 \right) \tilde{a}^{\dagger}_{\pmb{k}} \tilde{a}_{\pmb{k}},
\end{align}
where $\omega_0 \equiv |\gamma| \mu_0 H_0$ is the ferromagnetic resonance frequency ($\sim$ GHz) corresponding to the uniform ($\pmb{k} = \pmb{0}$) magnon mode, $a$ is the lattice constant, $S$ is the spin, $c_l$ is a factor that depends on the considered lattice, $\tilde{a}_{\pmb{k}}$ denotes the annihilation operator for a magnon with wavevector $\pmb{k}$. {The magnons here bear unit spin as each of them reduces the total spin in the ferromagnet by that amount~\cite{Kittel1963}.} The boundary conditions for small magnets result in a discrete magnon spectrum~\cite{Stancil2009}. {This leads to discrete allowed values of the wavevector $\pmb{k}$ leaving the Hamiltonian unchanged otherwise. Furthermore, $\pmb{k}$ then labels standing waves instead of traveling waves.} For typical values of $J$, spatial dimensions in the $\mu$m range result in the magnon energies differing by a few GHz. Hence, we consider only the $\pmb{k} = \pmb{0}$ mode henceforth, denoting $\tilde{a}_{\pmb{0}}$ simply as $\tilde{a}$. We may disregard the higher modes as we exploit coherent resonant interactions in this study.

As depicted in Fig.~\ref{fig:1}(a), the confined electron gas that becomes a spin qubit is interfaced directly with the ferromagnet to enable exchange-coupling~\cite{Takahashi2010,Bender2015,Kamra2016B,Trifunovic2013}:
\begin{align}\label{eq:hint0}
\tilde{H}_{\mathrm{int}} & =  J_{\mathrm{int}} \sum_{l} \tilde{\pmb{S}}_l \cdot \tilde{\pmb{s}}_l ,
\end{align}
where $J_{\mathrm{int}}$ parameterizes the interfacial exchange interaction, $\tilde{\pmb{s}}_l$ denotes the spin operator of the spin qubit electronic state at site $l$, and $l$ runs over the interfacial sites. In terms of the relevant eigenmodes, the interfacial interaction is simplified as:
\begin{align}\label{eq:hint1}
\tilde{H}_{\mathrm{int}} & = g \left( \tilde{a}^\dagger \tilde{\sigma}_{-} + \tilde{a} \tilde{\sigma}_{+} \right) ,
\end{align}
where $g = J_{\mathrm{int}} N_{\mathrm{int}} |\psi|^2 \sqrt{S/(2N_F)}$, with $N_{\mathrm{int}}$ the number of interfacial sites, $|\psi|^2$ the spin qubit electron probability averaged over the interface, and $N_F$ the total number of sites in the ferromagnet. $\tilde{\sigma}_{+,-} = (\tilde{\sigma}_{x} \pm i \tilde{\sigma}_y)/2$ excite or relax the spin qubit that is further described via:
\begin{align}\label{eq:qubitham1}
\tilde{H}_{\mathrm{q}} & = \frac{\omega_q}{2} \tilde{\sigma}_{z}.
\end{align} 
Thus, our total Hamiltonian becomes
\begin{align}\label{eq:totham1}
\tilde{H}_1 & = \tilde{H}_{\mathrm{F}} +  \tilde{H}_{\mathrm{q}} + \tilde{H}_{\mathrm{int}},
\end{align}
where $\tilde{H}_{\mathrm{F}} = \omega_0 \tilde{a}^{\dagger} \tilde{a}$ and the other contributions are given by Eqs.~(\ref{eq:hint1}) and (\ref{eq:qubitham1}). 

Our system thus realizes the Jaynes-Cummings Hamiltonian [Eq.~\eqref{eq:totham1}] that conserves the total number of excitations. This is a direct consequence of spin conservation afforded by the exchange-coupling in our system. A spin-1 magnon can be absorbed by a spin qubit flipping the latter from its spin $-1/2$ to $+1/2$ state. The same transition in the spin qubit, however, cannot emit a magnon. This is in contrast with the case of dipolar coupling between the spin qubit and the ferromagnet~\cite{Fukami2021,Akhiezer1968,Flebus2018,Du2017,Trifunovic2013}, which does not necessarily conserve spin. Further, as numerically estimated below, on account of exchange being a much stronger interaction, the effective coupling $g$ in our system can exceed the magnon frequency $\omega_0$ thereby covering the full coupling range from weak to deep-strong~\cite{Kockum2019,Casanova2010,Niemczyk2010}. Nonclassical behavior is typically manifested starting with ultrastrong couplings $g/\omega_0 > 0.1$~\cite{Forn2019,Ashhab2010,Kockum2019}.

We have considered the ferromagnet to be isotropic thus far. However, such films manifest a strong shape anisotropy, in addition to potential magnetocrystalline anisotropies~\cite{Akhiezer1968}. We now {account for these effects by including the single-ion anisotropy contribution parameterized via $K_{x,y,z}$:}
\begin{align}\nonumber
\tilde{H}_{\mathrm{an}} & = \sum_{i} K_x \left( \tilde{S}_{ix} \right)^2 + K_y \left( \tilde{S}_{iy} \right)^2 + K_z \left( \tilde{S}_{iz} \right)^2.
\end{align}
{Our assumed general form for the anisotropy allows us to capture all possible contributions to the uniform magnon mode Hamiltonian and provides design principles in choosing the right material. The specific cases of shape anisotropy~\cite{Kamra2016A,Kamra2016B} and magnetocrystalline anisotropy in specific materials~\cite{Zhang2019} are adequately captured by our general considerations, and have been detailed elsewhere~\cite{Kamra2016A,Kamra2016B,Zhang2019}. Retaining only the uniform mode, the anisotropy contribution above results in the following magnon Hamiltonian:}
\begin{align}\label{eq:sqham1}
\tilde{H}_{\mathrm{F}} & = A \tilde{a}^{\dagger} \tilde{a} + B \left( \tilde{a}^2 + \tilde{a}^{\dagger 2} \right),
\end{align}
with $A \equiv |\gamma|\mu_0 H_0 + K_x S + K_y S - 2 K_z S$ and $B \equiv S(K_x - K_y)/2$. {For typical physical systems, both $A$ and $B$ are in the GHz regime and are determined via the applied field and anisotropies as delineated by the expressions presented above.} The ensuing Hamiltonian Eq.~\eqref{eq:sqham1} possesses the squeezing terms $\propto B$ which, unlike in the case of light, result from the magnet trying to minimize its ground state energy while respecting the Heisenberg uncertainty principle~\cite{Kamra2020}. The new eigenmode, dubbed squeezed-magnon~\cite{Kamra2016A}, is obtained via a Bogoliubov transform $\tilde{a} = \cosh r \tilde{\alpha} + \sinh r \tilde{\alpha}^{\dagger}$ resulting in
\begin{align}\label{eq:sqham2}
\tilde{H}_{\mathrm{F}} & = \omega_0 \tilde{\alpha}^{\dagger} \tilde{\alpha},
\end{align}
where we continue to denote the eigenmode energy as $\omega_0$, which now becomes $\omega_0 = \sqrt{A^2 - 4 B^2}$. Further, the squeeze parameter $r$ is governed by the relation $\sinh r = -2B/\sqrt{(A + \omega_0)^2 - 4 B^2}$. {The ground state stability requires $\omega_0 > 0$ and $(A + \omega_0)^2 > 4 B^2$. Thus, while the physical system in question allows for $A$ and $B$ values outside this domain, our assumption of a uniformly ordered ground state becomes invalid in that case. We confine our analysis to the case of sufficiently large applied field $H_0$ such that the system harbors a uniformly ordered ground state. The limit of a divergent squeezing $r$ is nevertheless within the domain of applicability.} {A detailed analysis of squeezing resulting from shape anisotropy shows it to be a strong effect~\cite{Kamra2016A,Kamra2016B}, with $\sinh r$ of the order of unity for typical experiments. It can be much larger for small applied fields or when the magnet is close to a ground state instability or when a magnet with strong magnetocrystalline anisotropy is chosen. Further, the analysis above shows that breaking symmetry in the plane transverse to the equilibrium spin order yields the squeezing effect, while a uniaxial anisotropy does not contribute to it.} In the new eigenbasis, we obtain:
\begin{align}\label{eq:hint2}
\tilde{H}_{\mathrm{int}} & = g_R \left( \tilde{\alpha}^\dagger \tilde{\sigma}_{-} + \tilde{\alpha} \tilde{\sigma}_{+} \right) + g_{CR} \left( \tilde{\alpha}^\dagger \tilde{\sigma}_{+} + \tilde{\alpha} \tilde{\sigma}_{-} \right) ,
\end{align}
with $g_R = g \cosh r$ and $g_{CR} = g \sinh r$. The interaction now bears both rotating ($\propto g_{R}$) and counter-rotating ($\propto g_{CR}$) terms [Fig.~\ref{fig:1}(b)]. 

Our system can be analyzed in terms of two different bases: using spin-1 magnon (represented by $\tilde{a}$) or squeezed-magnon ($\tilde{\alpha}$). The latter is the eigenmode and is comprised of a superposition of odd magnon-number states [Fig.~\ref{fig:2}(a)]~\cite{Kamra2016A,Kamra2020,Nieto1997,Kral1990}. Since a spin-1 magnon is associated with the physical spin-flip in the magnet~\cite{Holstein1940}, the interaction Eq.~\eqref{eq:hint1} is still comprised of absorption and emission of magnons ($\tilde{a}$) accompanied by transitions in the qubit. On the other hand, in the eigenbasis, the qubit is now interacting with a new bosonic eigenmode - the squeezed-magnon ($\tilde{\alpha}$) via an interaction bearing rotating and CRTs [Eq.~\eqref{eq:hint2}]. Therefore, in the eigenbasis, our system accomplishes an anisotropic QRM~\cite{Xie2014,Xie2017} [Fig.~\ref{fig:1}(b)] - Eqs.~(\ref{eq:qubitham1}), (\ref{eq:totham1}), (\ref{eq:sqham2}), and (\ref{eq:hint2}). The squeeze parameter $r$, tunable via applied field and anisotropies~\footnote{The magnetocrystalline anisotropies can be tuned via strain, for example.}, further enhances the coupling strength and controls the relative importance of the rotating and CRTs: $g_R = g \cosh r$ and $g_{CR} = g \sinh r$. 

\section{1 magnonic mode coupled to 3 qubits}\label{sec:1mag3qubits}

\begin{figure}[tb]
	\begin{center}
		\includegraphics[width=85mm]{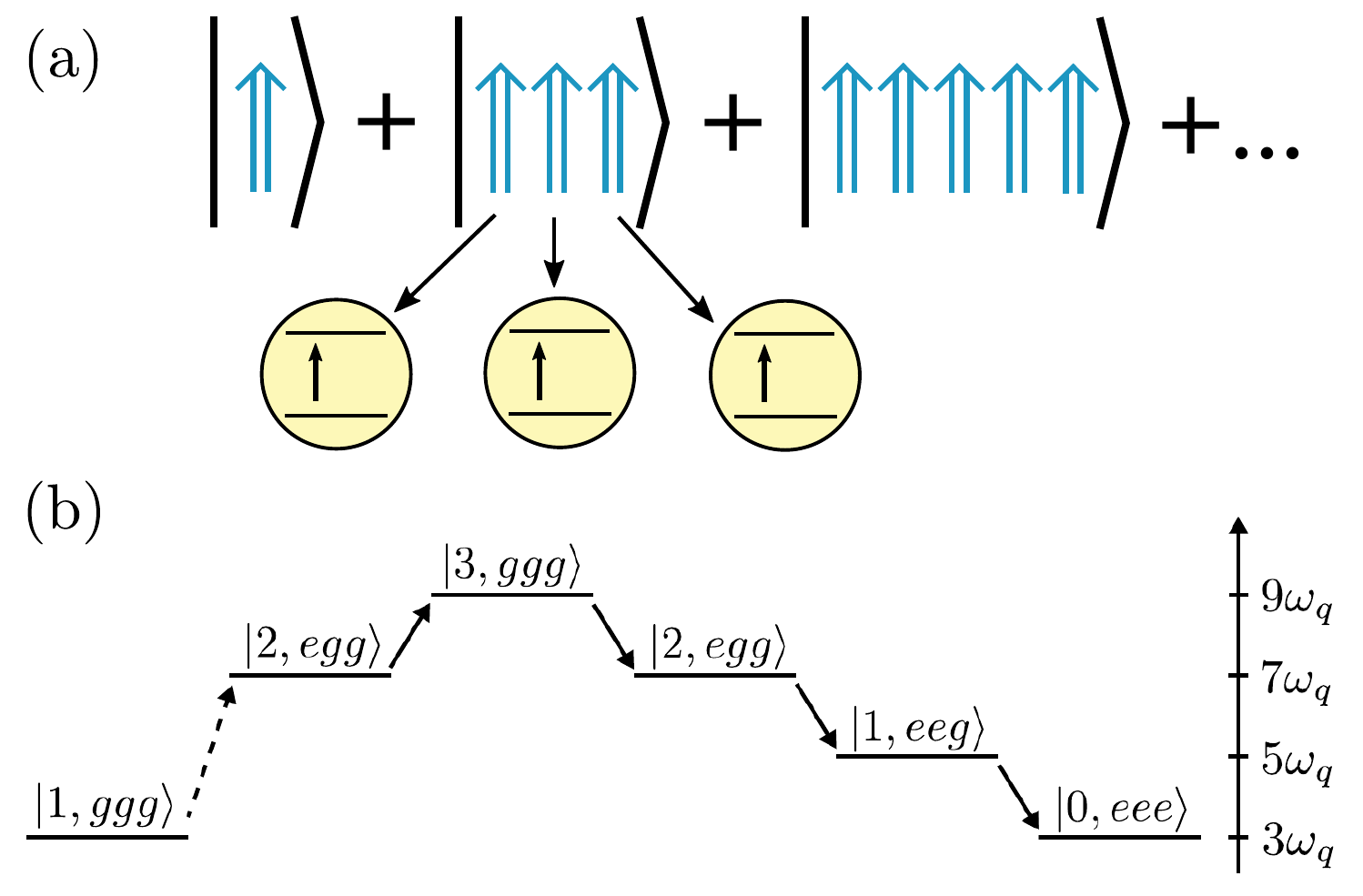}
		\caption{Schematic depiction of the transition $\ket{1,ggg} \to \ket{0,eee}$. (a) The squeezed-magnon is comprised of a superposition of odd magnon-number states. This composite nature enables its absorption by an odd number of qubits. We focus on the case of 3. (b) An example pathway that takes the system from bearing 1 squeezed-magnon and 3 ground-state qubits ($\ket{1,ggg}$) to $0$ squeezed-magnon and 3 excited qubits ($\ket{0,eee}$) via a series of virtual states. The first transition is effected by a CRT and is indicated via a dashed arrow. The right scale indicates the state energy, assuming $\omega_0 = 3 \omega_q$.}
		\label{fig:2}
	\end{center}
\end{figure}

We now exploit the squeezed and composite nature of the magnonic eigenmode in generating useful entangled states~\cite{Kamra2020}. As depicted in Fig.~\ref{fig:2}(a), the composite nature of the squeezed-magnon should enable joint excitation of an odd number of qubits. Considering the paramount importance of generating such 3-qubit GHZ states~\cite{Greenberger2007} for Shor's error correction code~\cite{Shor1995}, we consider 3 qubits coupled to the same squeezed-magnon eigenmode:
\begin{align}\label{eq:h3}
\tilde{H}_3 & = \tilde{H}_{\mathrm{F}} + \sum_{n = 1,2,3} \left( \tilde{H}_{\mathrm{q}}^n +  \tilde{H}_{\mathrm{int}}^n \right),
\end{align}
with individual contributions expressed via Eqs. (\ref{eq:qubitham1}), (\ref{eq:sqham2}), and (\ref{eq:hint2}). For simplicity, we assume the three qubits and their coupling with the magnet to be identical. The qualitative physics is unaffected by asymmetries among the 3 qubits, which are detailed in {appendix \ref{sec:pert}}. Henceforth, we analyze the problem in its eigenbasis employing methodology consistent with previous investigation of joint photon absorption~\cite{Garziano2016}.

We are interested in jointly exciting the three qubits using a single squeezed-magnon eigenmode: a transition denoted as $\ket{1,ggg} \to \ket{0,eee}$. To gain physical insight, we first analyze this transition within the perturbation-theory framework detailed in {appendix \ref{sec:pert}}. While the transition is not possible via a direct process [first order in the interaction Eq.~(\ref{eq:hint2})], it can be accomplished via a series of virtual states. As the transition requires an increase of the total excitation number by 2, at least one of the virtual processes should be effected via the CRTs, thus requiring nonzero squeezing $r$ in our system. The shortest path to effect the transition consists of three virtual processes, but its amplitude is canceled exactly by a complementary path, as detailed in appendix \ref{sec:pert}. Hence, the lowest nonvanishing order for accomplishing this transition is five with an example pathway depicted in Fig.~\ref{fig:2}(b)~\footnote{Since this path involves 4 rotating and 1 counter-rotating processes, its amplitude scales as $\sim g_{CR} g_{R}^4$.}. As detailed in {appendix \ref{sec:pert}}, several such paths contribute to the overall transition amplitude. The energy conservation requirement on the initial and final states necessitates $\omega_0 \approx 3 \omega_q$.

\begin{figure*}[tb]
	\begin{center}
		\includegraphics[width=180mm]{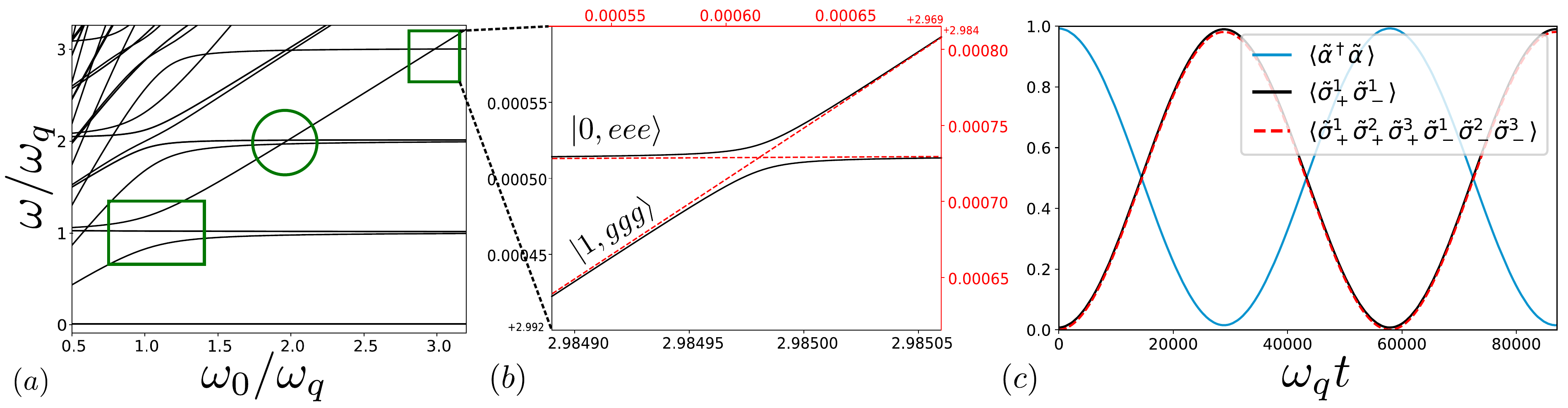}
		\caption{Numerically evaluated spectrum and dynamics of 3 qubits coupled to 1 magnonic mode [Eq.~\eqref{eq:h3}]. (a) Energy spectrum evaluated assuming $g_{R} = g_{CR} = 0.1 \omega_q$. The green rectangle encloses the typical one-excitation anticrossing ($\omega_0 \approx \omega_q$). The circle highlights crossings around $\omega_0 \approx 2\omega_q$ as only odd number of qubits can be excited [Fig.~\ref{fig:2}(a)]. The square emphasizes the weaker three-excitation anticrossing around $\omega_0 \approx 3\omega_q$ that results from finite squeezing and the resulting CRTs. (b) A zoom-in on the three-excitation anticrossing that stems from the transition depicted in Fig.~\ref{fig:2}. The red dashed lines depict the spectrum evaluated assuming $g_{CR} = 0$ leaving the rest unchanged. (c) Zero-detuning system dynamics around $\omega_0 \approx 3\omega_q$ with the initial state $\ket{1,ggg}$. The squeezed-magnon occupation (blue solid) and single-qubit excitation (black solid) manifest the typical Rabi oscillations. A nearly perfect overlap between single-qubit and three-qubit (red dashed) correlations confirms the joint nature of the three-qubit excitation in these Rabi oscillations.}
		\label{fig:3}
	\end{center}
\end{figure*}

Guided by intuition from the perturbative analysis, we now study the system [Eq.~\eqref{eq:h3}] numerically using the QuTiP package~\cite{QuTiP1,QuTiP2}. Unless stated otherwise {and for simplicity}, we employ $g_{R} = g_{CR} = 0.1 \omega_q$ in our analysis. A numerical diagonalization of the total Hamiltonian Eq.~\eqref{eq:h3} yields the energy spectrum as depicted in Fig.~\ref{fig:3}(a). {To understand it, let us first consider the simpler case of zero qubit-magnon coupling. In that case, the} spectrum should contain 8 ($2^3$) flat curves corresponding to the different excited qubits and zero squeezed-magnon occupation. Two triplets of these overlap resulting in 4 visually-distinct flat curves. The same 3-qubit spectrum combined with $N$ squeezed-magnons yields the same 4 visually-distinct curves, now with a slope of $N$. {Let us turn on the qubit-magnon coupling now.} For small but finite coupling considered in Fig.~\ref{fig:3}(a), we see the typical one-excitation Rabi splitting around $\omega_0 \approx \omega_q$ that results from a direct process. Around $\omega_0 \approx 2 \omega_q$, we see crossings between different levels~\footnote{The nature ``crossing'', as opposed to anticrossing, of these intersections has been verified carefully by evaluating the energy spectra around them with a very high precision.}. A coupling here is forbidden as only odd number of qubits can be excited by one squeezed-magnon [Fig.~\ref{fig:2}(a)]. An apparent crossing around $\omega_0 \approx 3 \omega_q$ is in fact an anticrossing manifesting a small Rabi splitting between the states $\ket{1,ggg}$ and $\ket{0,eee}$ [see Fig.~\ref{fig:3}(b)]. This is the transition of our interest and the effective coupling responsible for it can be expressed as:
\begin{align}\label{eq:heff}
\tilde{H}_{\mathrm{eff}} & = g_{\mathrm{eff}} \left( \ket{1,ggg}\bra{0,eee} + \ket{0,eee}\bra{1,ggg} \right),
\end{align}
where {$g_{\mathrm{eff}} = (g_{CR} g_R^4 - 0.3 g_{CR}^3 g_R^2)/\omega_q^4$ has been obtained by fitting (almost perfectly) its $g_{R,CR}$ dependence predicted by the perturbative analysis to the Rabi splittings obtained via numerical diagonalization. In carrying out this analysis, we numerically found the resonance condition which occurs around $\omega_0 \approx 3 \omega_q$ and evaluated the Rabi splitting. Hence, the expression for $g_{\mathrm{eff}}$ above is valid for $\omega_0 \approx 3 \omega_q$.} A comparison between squeezed-magnon occupation, single-qubit excitation, and three-qubit correlations plotted in Fig.~\ref{fig:3}(c) for Rabi oscillations around $\omega_0 \approx 3 \omega_q$ confirms the joint nature of the three-qubit excitation.


\section{Discussion}\label{sec:discussion}

Our system enables the transition $\ket{1,ggg} \to \ket{0,eee}$ with an effective coupling strength  $g_{\mathrm{eff}}$ [Eq.~(\ref{eq:heff})], or equivalently Rabi frequency, tunable via the magnon-squeezing: $g_{CR} = g 
\sinh r$. Bringing the system in resonance to enable Rabi oscillation for a fraction of the cycle can be exploited in robustly generating 3-qubit GHZ and related entangled states: $(\ket{ggg} \pm \ket{eee})/\sqrt{2}$. A convenient generation of these is central to Shor's error correction code~\cite{Shor1995} and thus, of great value in achieving fault tolerant quantum computing. {Such 3-qubit entangled states can be generated on contemporary quantum computers via sequential one- and two-qubit gate operations~\cite{Neeley2010,DiCarlo2010,Reed2012}. In theory, and for ideal gate operations, our suggested method appears to not offer any advantage over such sequential gate operations executed on state-of-the-art quantum computers. However, each two-qubit gate operation entails applying an exact pulse that in turn depends on the qubit frequencies and their coupling to the bosonic mode. Further, such sequential operations necessarily create an asymmetry between the three qubits since one of them needs to addressed in the end. In the presence of decoherence, this can compromise the quality of the GHZ states achieved in practice. Finally, sequential operations are bound to take a longer time in generating the desired GHZ state, which reduces the time available for other computations given that decoherence limits the total time available.} In contrast, capitalizing on energy and spin conservation, our proposed {single-pulse} method is intrinsically robust against any qubit asymmetries and perfectly synchronizes excitation of the 3 qubits. {This resilience of our suggested method comes as there is a unique resonance condition around $\omega_0 \approx \omega_{q1} + \omega_{q2} + \omega_{q3}$ for the single pulse needed. Since the three qubits need to absorb the energy of one squeezed-magnon together, their GHZ state generation is automatically synchronized.}

Being a fifth-order process, $g_{\mathrm{eff}}$ was evaluated to be small for the parameters employed in our analysis above ($g_{R} = g_{CR} = 0.1 \omega_q$). However, notwithstanding our choice of parameters motivated by a comparison with perturbation theory, the proposed system can achieve very high bare couplings $g$ [Eq.~\eqref{eq:hint2}] ($g_{R},g_{CR} > \omega_q$), such that the higher-order processes are not diminished and $g_{\mathrm{eff}}$ becomes large. {An increase in the coupling strength and the relevance of higher-order processes, however, has its trade-offs. While some of these higher-order processes merely renormalize the qubit and magnon frequencies thereby not affecting the phenomena discussed herein, others can bring the independent existence of the magnon and qubit subsystems into question. Thus, depending on the desired application, an optimal value for $g_{\mathrm{eff}}$ needs to be chosen. The key benefit of the proposed system is the wide range of $g_{\mathrm{eff}}$ that it admits.} Spin pumping experiments yield interfacial exchange couplings [Eq.~\eqref{eq:hint0}] of $J_{\mathrm{int}} \approx 10$ meV between various (insulating) magnets and adjacent metals~\cite{Kajiwara2010,Czeschka2011,Weiler2013}. Assuming the qubit wavefunction to be localized in 5 monolayers below the equally thin ferromagnet and an interface comprised of 100 sites, we obtain the bare coupling rate [Eq.~\eqref{eq:hint1}] $g \approx 0.005 J_{\mathrm{int}} \approx 80$~GHz, significantly larger than typical spin qubit and uniform magnon mode frequencies. 

{In general, one can design a system (e.g., by choosing the ferromagnet thickness) to bear a desired bare coupling and exploit the squeezing-mediated tunability insitu. The latter effect, although an interesting and useful property of the system, may not be needed in the specific application given that deep-strong coupling could be achieved without the enhancement. In particular, our example of choice - the generation of GHZ states, need not exploit this enhancement effect.}

Our proposal of leveraging the intrinsic magnon-squeezing in generating entanglement via a coherent process is complementary to previous incoherent interaction-based proposals~\cite{Trifunovic2013,Kamra2019,Zou2020,Yuan2021}. The latter typically necessitate diabatic decoupling of qubits from the magnet after achieving an entangled state. Our proposal thus uncovers an unexplored and experimentally-favorable avenue for exploiting the squeezing intrinsic to magnets.


\section{Summary}\label{sec:summary}

We have demonstrated the magnon--spin-qubit ensemble to realize the anisotropic quantum Rabi model with coupling strengths feasible in the deep-strong regime. This system is shown to capitalize on various unique features of squeezed-magnons hosted by magnets. These include squeezing-mediated coupling enhancement, tunable anisotropy of the Rabi model, and a convenient synchronous entanglement of 3 qubits. Thus, the magnon--spin-qubit ensemble provides a promising platform for investigating phenomena beyond the ultrastrong regime and implementing error correction codes.


\section*{Acknowledgment}

We thank Wolfgang Belzig, Tim Ludwig, and Rembert Duine for valuable discussions. We acknowledge financial support from the Research Council of Norway through its Centers of Excellence funding scheme, project 262633, ``QuSpin'', and the Spanish Ministry for Science and Innovation -- AEI Grant CEX2018-000805-M (through the ``Maria
de Maeztu'' Programme for Units of Excellence in R\&D).

\appendix

\section{System Hamiltonian}\label{sec:ham}

In this section, we derive the Hamiltonian describing our magnon/spin-qubit ensemble. First, starting with the ferromagnetic spin Hamiltonian, we obtain the description of the magnonic mode. Then, we specify the spin-qubit. Finally, we derive the interfacial exchange-mediated interaction between the two subsystems.

\subsection{Magnonic mode}
Taking into account Zeeman energy, ferromagnetic exchange, and a general anisotropy, the ferromagnet is described via the spin Hamiltonian:
\begin{align}\label{eq:hspin}
\tilde{H}_{\mathrm{F}}  = & |\gamma| \mu_0 H_0 \sum_{i} \tilde{S}_{iz} - J \sum_{\left\langle i,j \right\rangle}  \tilde{\pmb{S}}_i \cdot \tilde{\pmb{S}}_j  \nonumber \\ 
   & + \sum_{i} \left[ K_x \left( \tilde{S}_{ix} \right)^2 + K_y \left( \tilde{S}_{iy} \right)^2 + K_z \left( \tilde{S}_{iz} \right)^2 \right],
\end{align}
where the applied magnetic field is $H_0 \hat{\pmb{z}}$, $\gamma~(<0)$ is the gyromagnetic ratio, $J~(>0)$ is the exchange energy, $\left\langle i,j \right\rangle$ denotes sum over nearest neighbors, and $K_{x,y,z}$ parameterize the magnetic anisotropy. While the anisotropy may arise due to dipolar interactions or magnetocrystalline single-ion anisotropies, our assumed general form encompasses all such symmetry-allowed contributions that can contribute to determining the uniform $\pmb{k} = \pmb{0}$ magnon mode~\cite{Kamra2016B}. 

Assuming the Zeeman energy to dominate over anisotropy, we consider all spins to point along $- \hat{\pmb{z}}$ in the magnetic ground state. We may express the spin Hamiltonian Eq.~\eqref{eq:hspin} in terms of bosonic magnons via the Holstein-Primakoff transformation~\cite{Holstein1940} corresponding to our spin ground state:
\begin{align}
\tilde{S}_{j+} & = \sqrt{2S} ~ \tilde{a}_j^\dagger, \label{eq:hp1} \\
\tilde{S}_{j-} & = \sqrt{2S} ~ \tilde{a}_j, \\
\tilde{S}_{jz} & = -S + \tilde{a}_j^\dagger \tilde{a}_j,
\end{align}
where $\tilde{S}_{j\pm} \equiv \tilde{S}_{jx} \pm i \tilde{S}_{jy}$, $\tilde{a}_j$ is the magnon annihilation operator at position $j$, and $S$ is the spin magnitude. In addition, we need the Fourier relations:
\begin{align}
\tilde{a}_{j} & = \frac{1}{\sqrt{N_F}} \sum_{\pmb{k}} \tilde{a}_{\pmb{k}} ~ e^{- i \pmb{k} \cdot \pmb{r}_{j}}, \\
\tilde{a}_{\pmb{k}} & = \frac{1}{\sqrt{N_F}} \sum_{j} \tilde{a}_{j} ~ e^{i \pmb{k} \cdot \pmb{r}_{j}}, \label{eq:ft2}
\end{align}
where $N_F$ is the total number of sites in the ferromagnet and $\tilde{a}_{\pmb{k}}$ is the annihilation operator for the magnon mode with wavevector $\pmb{k}$. Employing these Holstein-Primakoff and Fourier transformations in Eq.~\eqref{eq:hspin}, we obtain the magnonic Hamiltonian:
\begin{align}\label{eq:hmag}
\tilde{H}_{\mathrm{F}} & = \mathrm{const.} + \sum_{\pmb{k}} \left[ A_{\pmb{k}} \tilde{a}_{\pmb{k}}^\dagger \tilde{a}_{\pmb{k}} + B_{\pmb{k}} \left( \tilde{a}_{\pmb{k}}^\dagger \tilde{a}_{-\pmb{k}}^\dagger + \tilde{a}_{\pmb{k}} \tilde{a}_{-\pmb{k}} \right)   \right],
\end{align}
with $A_{\pmb{k}} \equiv |\gamma|\mu_0 H_0 + K_x S + K_y S - 2 K_z S + 4 J S \left[ 3 - \left( \cos k_x a + \cos k_y a + \cos k_z a \right) \right]$ and $B_{\pmb{k}} \equiv S(K_x - K_y)/2$. In obtaining the exchange contribution to $A_{\pmb{k}}$, we have assumed a simple cubic lattice with lattice constant $a$. In the long wavelength limit i.e., $a k_{x,y,z} \ll 1$, the cosines can be approximated by parabolas. 

As discussed in the main text, we retain only the uniform mode corresponding to $\pmb{k} = \pmb{0}$ in our consideration of the magnon/spin-qubit system. This is justifiable because for small dimensions of the magnet considered herein, the allowed wavevectors $\pmb{k}$ correspond to magnon energies separated from the lowest uniform mode (with an energy of a few GHz) by at least several GHz. Thus, we may disregard such high-energy modes when considering coherent resonant interactions, as we do in this work. Further diagonalization of Eq.~\eqref{eq:hmag}, considering only the uniform mode, via Bogoliubov transformation has been described in the main text.

\subsection{Spin-qubit}
We consider our spin-qubit to be comprised by a confined electronic orbital that admits spin-up and -down states. Considering a lifting of the spin-degeneracy by, for example, an applied magnetic field, the spin-qubit Hamiltoniam may be expressed as:
\begin{align}\label{eq:hq1}
\tilde{H}_{\mathrm{q}} & =  \mathrm{const.} + \frac{\omega_q}{2} \left( \tilde{c}_{\uparrow}^\dagger \tilde{c}_{\uparrow} - \tilde{c}_{\downarrow}^\dagger \tilde{c}_{\downarrow} \right),
\end{align}
where, considering a negative gyromagnetic ratio and applied magnetic field along $\hat{\pmb{z}}$, $\omega_q~(>0)$ is the qubit splitting. We further introduce the notation:
\begin{align}\label{eq:sigdef}
\tilde{\sigma}_{z} & \equiv \begin{pmatrix}
\tilde{c}_{\uparrow}^\dagger & \tilde{c}_{\downarrow}^\dagger
\end{pmatrix} \begin{pmatrix}
1 & 0 \\ 0 & -1
\end{pmatrix} \begin{pmatrix}
\tilde{c}_{\uparrow} \\ \tilde{c}_{\downarrow}
\end{pmatrix} \ \equiv \ \underline{\tilde{c}}^\dagger ~ \underline{\sigma}_z ~ \underline{\tilde{c}},
\end{align}
where an underline identifies a matrix. With this notation and dropping the spin-independent constant, the spin-qubit Hamiltonian is expressed as:
\begin{align}\label{eq:wqfin}
\tilde{H}_{\mathrm{q}} & = \frac{\omega_q}{2} \tilde{\sigma}_z.
\end{align}
With the notation defined by Eq.~\eqref{eq:sigdef}, $\tilde{\sigma}_+ \equiv (\tilde{\sigma}_x + i \tilde{\sigma}_y)/2$ becomes the qubit excitation operator, while $\tilde{\sigma}_- \equiv (\tilde{\sigma}_x - i \tilde{\sigma}_y)/2$ is the qubit relaxation operator.

\subsection{Exchange coupling}
The magnon/spin-qubit are considered to be coupled via interfacial exchange interaction parameterized via $J_{\mathrm{int}}$~\cite{Takahashi2010,Bender2015,Kamra2016B}:
\begin{align}\label{eq:hint0app}
\tilde{H}_{\mathrm{int}} & =  J_{\mathrm{int}} \sum_{l} \tilde{\pmb{S}}_l \cdot \tilde{\pmb{s}}_l ,
\end{align}
where $l$ labels the interfacial sites, $\tilde{\pmb{S}}$ denotes the ferromagnetic spin operator, and $\tilde{\pmb{s}}$ represents the spin of the electronic states that comprise the qubit. We wish to express the interfacial Hamiltonian Eq.~\eqref{eq:hint0app} in terms of the magnon and qubit operators. To this end, $\tilde{\pmb{S}}_l$ can be expressed via magnon operators using the Holstein-Primakoff and Fourier transforms [Eqs.\eqref{eq:hp1} - \eqref{eq:ft2}] already described above. We now discuss the representation of $\tilde{\pmb{s}}_l$ in terms of the qubit operators $\tilde{\sigma}_{x,y,z}$ [Eq.~\eqref{eq:sigdef}].

Following quantum field theory notation for discrete sites, the spin operator at a given position $\pmb{r}$ can be expressed in terms of ladder operators at the same position:
\begin{align}\label{eq:sq1}
\tilde{\pmb{s}}(\pmb{r}) & = \frac{1}{2} \sum_{s,s\prime = \uparrow,\downarrow} \tilde{\Psi}_{s}^\dagger (\pmb{r}) \pmb{\sigma}_{s s^\prime}  \tilde{\Psi}_{s^\prime} (\pmb{r}),
\end{align}
where $\underline{\pmb{\sigma}} = \underline{\sigma}_x \hat{\pmb{x}} + \underline{\sigma}_y \hat{\pmb{y}} + \underline{\sigma}_z \hat{\pmb{z}}$ with $\underline{\sigma}_{x,y,z}$ the Pauli matrices. The local ladder operators can further be represented in terms of the complete set of eigenstates labeled via orbital index $t$:
\begin{align}
\tilde{\Psi}_{s} (\pmb{r}) & = \sum_{t} \psi_t(\pmb{r}) \tilde{c}_{ts},
\end{align}
where $\psi(\pmb{r})$ is the spatial wavefunction of the different orbitals and $\tilde{c}_{ts}$ are the ladder operators for each spin-resolved orbital. Employing this relation, Eq.~\eqref{eq:sq1} becomes:
\begin{align}
\tilde{\pmb{s}}(\pmb{r}) & = \frac{1}{2} \sum_{s,s\prime,t,t^\prime} \psi_t^*(\pmb{r}) \psi_{t^\prime}(\pmb{r}) \pmb{\sigma}_{s s^\prime} \tilde{c}_{ts}^\dagger \tilde{c}_{t^\prime s^\prime}.
\end{align}
Since for our spin-qubit we are interested in only one of the complete set of orbitals, we allow only 1 value of $t$ and thus drop the index $t$ in consistence with our previous considerations Eq.~\eqref{eq:hq1}:
\begin{align}
\tilde{\pmb{s}}(\pmb{r}) & = \frac{1}{2} \sum_{s,s\prime} \left|\psi(\pmb{r}) \right|^2 \pmb{\sigma}_{s s^\prime} \tilde{c}_{s}^\dagger \tilde{c}_{s^\prime}, \\
& = \frac{\left|\psi(\pmb{r}) \right|^2}{2} ~ \underline{\tilde{c}}^\dagger ~ \underline{\pmb{\sigma}} ~ \underline{\tilde{c}}, \\
\implies \tilde{\pmb{s}}_l & = \frac{\left|\psi_l \right|^2}{2} ~ \underline{\tilde{c}}^\dagger ~ \underline{\pmb{\sigma}} ~ \underline{\tilde{c}}, \label{eq:slfin}
\end{align}
where $\psi_l$ is the wavefunction amplitude of the qubit orbital at position $l$.

The interfacial interaction Eq.~\eqref{eq:hint0app} is now simplified as:
\begin{align}
\tilde{H}_{\mathrm{int}} & = J_{\mathrm{int}} \sum_{l} \left[ \tilde{S}_{lz} \tilde{s}_{lz} + \frac{1}{2} \left( \tilde{S}_{l+}\tilde{s}_{l-} + \tilde{S}_{l-}\tilde{s}_{l+} \right)  \right] ,  
\end{align}
where $\tilde{S}_{l\pm} \equiv \tilde{S}_{lx} \pm i \tilde{S}_{ly}$ and $\tilde{s}_{l\pm} \equiv \tilde{s}_{lx} \pm i \tilde{s}_{ly}$. Employing Eq.~\eqref{eq:slfin} together with Eqs.\eqref{eq:hp1} - \eqref{eq:ft2} and retaining only the uniform magnon mode, the interfacial Hamiltonian is simplified to include two contributions:
\begin{align}
\tilde{H}_{\mathrm{int}} & = \tilde{H}_{\mathrm{int1}} + \tilde{H}_{\mathrm{int2}}. 
\end{align}
The first contribution is our desired magnon/spin-qubit exchange coupling:
\begin{align}
\tilde{H}_{\mathrm{int1}} & = J_{\mathrm{int}} N_{\mathrm{int}} |\psi|^2 \sqrt{\frac{S}{2N_F}} \left( \tilde{a}^\dagger \tilde{\sigma}_{-} + \tilde{a} \tilde{\sigma}_{+} \right) ,
\end{align}
where $N_{\mathrm{int}}$ is the number of interfacial sites and $|\psi|^2 \equiv \left( \sum_l |\psi_l|^2 \right) / N_{\mathrm{int}}$ is the qubit electronic state wavefunction averaged over the interface. The second contribution:
\begin{align}
\tilde{H}_{\mathrm{int2}} & = - \frac{S J_{\mathrm{int}} N_{\mathrm{int}} |\psi|^2}{2} ~ \tilde{\sigma}_z
\end{align}
renormalizes the spin-qubit energy and can be absorbed into $\omega_q$ [Eq.~\eqref{eq:wqfin}].


\section{Deriving an expression for the effective coupling with perturbation theory}\label{sec:pert}
Here we look at the Hamiltonian describing 3 qubits coupled to the same squeezed-magnon eigenmode, as described in the main text. We assume the interaction terms, $\tilde{H}_{\mathrm{int}}^n$, to be small compared to the rest of the Hamiltonian, $\tilde{H}_0 = \omega_0\tilde{\alpha}^\dagger\tilde{\alpha} + \sum_{n=1,2,3} \frac{\omega_{qn}}{2}\tilde{\sigma}_{z}^n$, and calculate the effective coupling $g_{\mathrm{eff}}$ between the two states $\ket{1,ggg}$ and $\ket{0,eee}$ using perturbation theory. The interaction term, $\tilde{H}_{\mathrm{int}}^n$, is given by:
\begin{gather}\label{eq:hc}
\tilde{H}_{\mathrm{int}}^n =  g_{Rn} \left(\tilde{\alpha}^\dagger\tilde{\sigma}_{-}^n +\tilde{\alpha}\tilde{\sigma}_{+}^n \right) + g_{CRn} \left(\tilde{\alpha}^\dagger\tilde{\sigma}_{+}^n + \tilde{\alpha}\tilde{\sigma}_{-}^n\right).
\end{gather}

The relevant virtual processes will be shown as paths from $\ket{1,ggg}$ (blue) to $\ket{0,eee}$ (red) on a grid of "number of magnon excitations" and "number of qubit excitations". The rotating term  (drawn as a full line) will keep the number of total excitations constant while the counter-rotating term (drawn as a dotted line) will change the total number of excitations by two. The detailed expressions for each diagram will be calculated using the diagrammatic approach from Ref. \cite{Salzman1968}. 

\subsection{Third-order perturbation theory}
\begin{figure}[tbh]
	\centering
	\includegraphics[width = 0.45\textwidth]{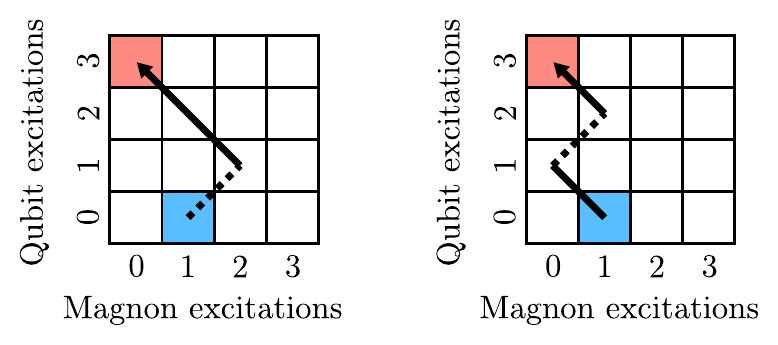}
	\caption{Diagrams connecting $\ket{1,ggg}$ (blue) and $\ket{0,eee}$ (red) via virtual transitions to third order. Counter-rotating processes are represented by dashed lines.}
	\label{fig:thirdOrder}
\end{figure}

We start by applying perturbation theory to third order. The two third-order diagrams are shown in Figure \ref{fig:thirdOrder}. For general qubits, these two diagrams result in the effective coupling:
\begin{widetext}
\begin{equation} \label{eq:generalThirdOrder}
g_{\mathrm{eff}}^{(3)} = \sum_{\substack{i,j,k \\i\neq j \neq k \neq i }} \left[\frac{2g_{CRi}g_{Rj}g_{Rk}}{(-\omega_0 - \omega_{qi})(-\omega_{qi} - \omega_{qj})} + \frac{g_{Ri}g_{CRj}g_{Rk}}{(\omega_0 - \omega_{qi})(-\omega_{qi} - \omega_{qj})}\right],
\end{equation}
\end{widetext}
where the sum is over all qubit permutations.

If we assume that the qubits are identical ($g_{CRi} = g_{CR}, g_{Ri} = g_{R}, \omega_{qi} = \omega_q$), all qubit permutations are equivalent and the sum can be carried out by counting qubit permutations:
\begin{equation}\label{eq:thirdOrder}
g_{\mathrm{eff}}^{(3)} = 3{g_{R}}^2g_{CR}\frac{3\omega_q-\omega_0}{\omega_q\left({\omega_0}^2-{\omega_q}^2\right)}.
\end{equation}
As we can see, the two paths cancel at resonance, $\omega_0 = 3\omega_q$. Moreover, it can be shown from equation \eqref{eq:generalThirdOrder} that the third-order term cancels when $\omega_0 = \sum_i \omega_{qi}$. The pure third-order perturbation theory result is therefore zero. 

\subsection{Fifth-order perturbation theory}
\begin{figure}[tbh]
	\centering
	\includegraphics[width = 0.45\textwidth]{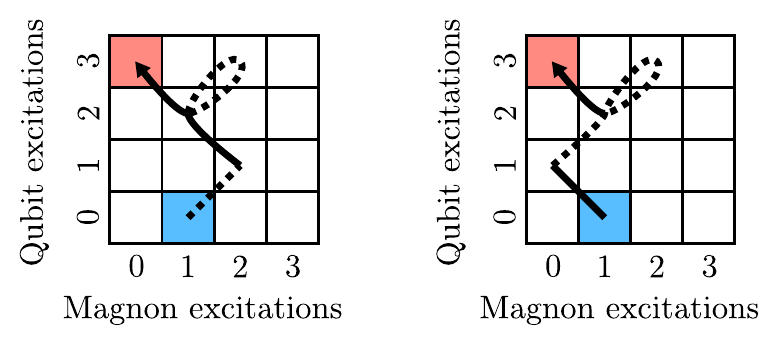}
	\caption{Example of fifth-order diagrams which cancel if $\omega_0 =3\omega_{q}$. Pairs of 
		the two third-order diagrams with an additional loop on a shared vertex that is not the initial vertex fully cancel at resonance. Counter-rotating terms are represented by dashed lines.}
	\label{fig:fifthOrderCancel}
\end{figure}

Since the third-order result is zero and there are no fourth-order paths, we move on to fifth order by drawing all fifth-order paths from the initial state $\ket{1,ggg}$ (blue) to the state $\ket{0,eee}$ (red). We use the result that the third-order term cancels at resonance to note that pairs of diagrams like the ones in Figure \ref{fig:fifthOrderCancel}, i.e. the two third-order diagrams with an additional loop on a shared vertex that is not the initial vertex, also fully cancel at resonance. 

\begin{figure*}[tbh]
	\centering
	\includegraphics[width = 130mm]{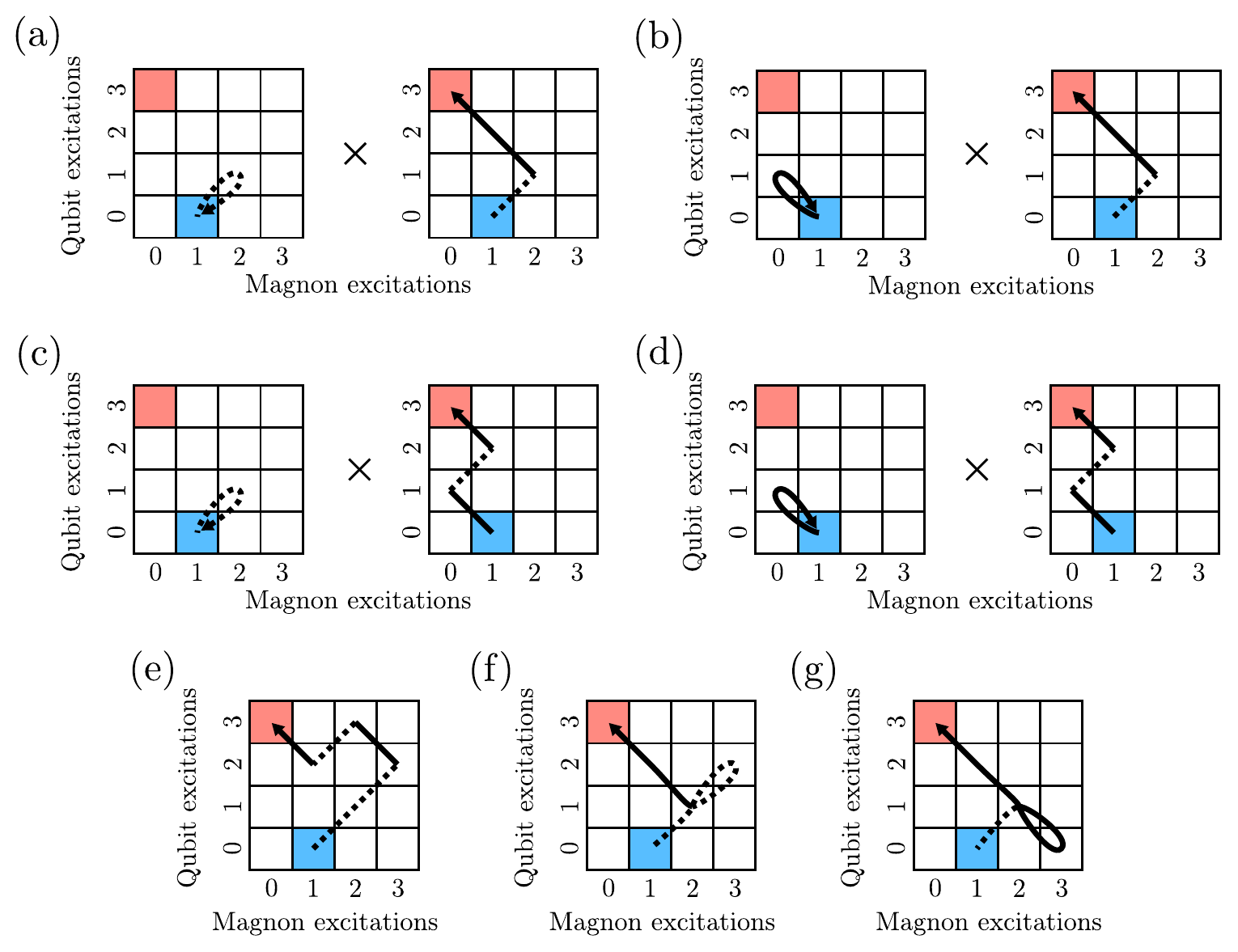}
	\caption{Relevant diagrams that connect $\ket{1,ggg}$ (blue) and $\ket{0,eee}$ (red) via virtual transitions to fifth order. Counter-rotating terms are represented by dashed lines.}
	\label{fig:fifthOrder}
\end{figure*}

\begin{widetext}
All remaining diagrams are shown in Figure \ref{fig:fifthOrder}. Diagrams (a) and (c) cancel partially, but not fully and give the contribution:
\begin{equation}
g_{\mathrm{eff}}^{(5a)} + g_{\mathrm{eff}}^{(5c)} =  \sum_{\substack{i,j,k,l \\j\neq k \neq l \neq j }} \left[\left(\frac{2{g_{CRi}}^2}{(-\omega_0 -\omega_{qi})}\right)\left(\frac{2g_{CRj}g_{Rk}g_{Rl}}{(-\omega_0 - \omega_{qi})^2(-\omega_{qi} - \omega_{qj})} + \frac{g_{Rj}g_{CRk}g_{Rl}}{(\omega_0 - \omega_{qi})^2(-\omega_{qi} - \omega_{qj})}\right)\right].
\end{equation}
Similarly for (b) and (d):
\begin{equation}
g_{\mathrm{eff}}^{(5b)} + g_{\mathrm{eff}}^{(5d)} =  \sum_{\substack{i,j,k,l \\j\neq k \neq l \neq j }} \left[\left(\frac{{g_{Ri}}^2}{(\omega_0 -\omega_{qi})}\right)\left(\frac{2g_{CRj}g_{Rk}g_{Rl}}{(-\omega_0 - \omega_{qi})^2(-\omega_{qi} - \omega_{qj})} + \frac{g_{Rj}g_{CRk}g_{Rl}}{(\omega_0 - \omega_{qi})^2(-\omega_{qi} - \omega_{qj})}\right)\right].\end{equation}
The contribution from diagram (e), (f) and (g):
\begin{equation}
g_{\mathrm{eff}}^{(5e)} = \sum_{\substack{i,j,k,l \\i\neq j \neq k \neq i }} \frac{6g_{CRi}g_{CRj}g_{Rk}g_{CRl}g_{Rl}}{(-\omega_0 - \omega_{qi})(-2\omega_0 - \omega_{qi}- \omega_{qj})(-\omega_0 -\sum_n \omega_{qn})(\omega_{ql}-\sum_n \omega_{qn} )}.
\end{equation}
\begin{equation}
g_{\mathrm{eff}}^{(5f)} = \sum_{\substack{i,j,k,l,m,n \\i\neq j \neq k \neq i\\ l = i,j\\m = l,k \\ n \neq m \quad n = l,k }} \frac{6g_{CRi}g_{CRj}g_{CRl}g_{Rm}g_{Rn}}{(-\omega_0 - \omega_{qi})(-2\omega_0 - \omega_{qi}- \omega_{qj})(-\omega_0 +\omega_{ql}+\omega_{qk}-\sum_p \omega_{qp})( \omega_{qn}-\sum_p \omega_{qp} )}.
\end{equation}
\begin{equation}
g_{\mathrm{eff}}^{(5g)} = \sum_{\substack{i,j,k,l \\j\neq k \neq l \neq j }} \frac{6g_{CRi}g_{Ri}g_{Rj}g_{Rk}g_{Rl}}{(-\omega_0 - \omega_{qi})(-2\omega_0)(-\omega_0 - \omega_{qj})(- \omega_{qj} - \omega_{qk})}.
\end{equation}
\end{widetext}

If we now assume that we are at resonance and that the qubits are identical ($g_{CRi} = g_{CR}, g_{Ri} = g_{R}, \omega_{qi} = \omega_q$ and $\omega_0 = 3\omega_q$), the sums can again be carried out by counting qubit permutations. The total effective coupling to fifth order is then:
\begin{equation}\label{eq:fifthOrderCoupling}
g_{\mathrm{eff}}^{(5)} = -\frac{9 \left(3 {g_{CR}}^3 {g_{R}}^2-8 g_{CR} {g_{R}}^4\right)}{32 {\omega_q}^4}.
\end{equation}

\subsection{Additional corrections}
As we have seen, the third-order contribution to the effective coupling is zero when $\omega_0 = 3\omega_q$. However, if we are interested in the details of the (anti-)crossing, there is an additional detail we need to consider. The perturbation causes the energy levels to shift, which causes the (anti-)crossing to take place a small shift away from $\omega_0 = 3\omega_q$. 

By applying second order perturbation theory (at $\omega_0 = 3\omega_q$) to the energies of the two relevant states, we get that the crossing will take place at:
\begin{equation}
\omega_0  = 3 \omega_{q} + \frac{3{g_{CR}}^2}{2\omega_q} - \frac{3{g_{R}}^2}{\omega_q}  .
\end{equation}
Inserting this into the third-order effective coupling, Eq. \eqref{eq:thirdOrder}, and keeping terms of up to fifth order in $g_{CR/R}$, leaves us with: \footnote{The effective coupling from Eq. \eqref{eq:fifthOrderCoupling} and Eq. \eqref{thirdOrderShiftedCoupling} can be tuned to zero, both separately (at $g_{CR} = \pm\sqrt{\frac{8}{3}}g_R$ and $g_{CR} =\pm \sqrt{2}g_R$ respectively) as well as the sum of the two ($g_{CR} = \pm 2g_R$).}

\begin{equation}\label{thirdOrderShiftedCoupling}
g_{\mathrm{eff}}^{(3)\prime} = \frac{9\left({g_{CR}}^3{g_{R}}^2-2 g_{CR}{g_{R}}^4\right)}{16 {\omega_q}^4},
\end{equation}
where the prime indicates that the effective coupling is evaluated at the (anti-)crossing.

\bibliography{ARabi}

\end{document}